\documentclass[conference]{IEEEtran}
\IEEEoverridecommandlockouts
\usepackage{cite}
\usepackage{amsmath,amssymb,amsfonts}
\usepackage{graphicx}
\DeclareGraphicsExtensions{.pdf,.png,.jpg,.jpeg}
\usepackage{textcomp}
\usepackage{xcolor}
\usepackage{booktabs}
\usepackage{tabularx}
\usepackage{array}
\usepackage{url}
\usepackage{stfloats}
\usepackage{caption}
\usepackage{balance}
\begin{document}

\title{Specificity- and Calibration-Aware Breast Ultrasound Segmentation via Entropy-Guided Boundary Supervision}

\author{

\IEEEauthorblockN{1\textsuperscript{st}Manar Alsaid}
\IEEEauthorblockA{Department of Computer Science\\
East Texas A\&M University\\
Commerce, USA\\
manar.alsaid@etamu.edu}
\and
\IEEEauthorblockN{1\textsuperscript{st}Mandip Shrestha}
\IEEEauthorblockA{Department of Computer Science\\
East Texas A\&M University\\
Commerce, USA\\
mshrestha18@leomail.tamuc.edu}
\and 
\IEEEauthorblockN{2\textsuperscript{nd}Mohammad Abbas}
\IEEEauthorblockA{
Texas Woman's University \\
Denton, USA\\
Abbashj@hotmail.com}

}

\maketitle

\begin{abstract}
Lesion segmentation in breast ultrasound presents two closely related challenges. In images containing lesions, speckle noise, low tissue contrast, and posterior acoustic shadowing frequently cause boundary leakage and incomplete contour delineation. In images without lesions, those same artifacts can produce highly confident false-positive activations in regions whose appearance resembles solid lesion tissue. This study addresses both failure modes through a single modification to the training objective. Rather than assigning equal weight to every boundary pixel, the proposed loss scales contour penalties according to per-pixel predictive entropy and the ground-truth boundary map. Gradient emphasis therefore concentrates on lesion margin locations for which the network remains genuinely uncertain. The proposed loss was evaluated on the BUSI dataset through a controlled ablation study against two baselines trained under identical conditions: a model without boundary supervision and a model with uniformly weighted boundary binary cross-entropy. Across the 97 lesion-containing test images, mean Dice scores were statistically indistinguishable between the proposed method and the no-boundary baseline (0.7624 versus 0.7616, paired Wilcoxon p = 0.27), indicating that lesion segmentation quality is preserved rather than improved. The primary effect appears in specificity. False-positive activations on the 20 no-lesion test images decreased from 14 of 20 for the no-boundary model and 19 of 20 for the standard boundary model to 5 of 20 with the proposed approach, with McNemar exact test p-values of 0.012 and 0.0005, respectively. Non-overlapping Wilson 95\% confidence intervals (25\% [11.2, 46.9] versus 70\% [48.1, 85.5] and 95\% [76.4, 99.1]) confirm that this specificity difference is both statistically significant and practically substantial. A post-hoc spatial temperature scaling procedure further reduced expected calibration error by more than half, from 0.0201 to 0.0095, without altering the thresholded segmentation masks. Taken together, these findings position entropy-guided boundary supervision and spatial calibration as training-level and inference-level refinements, respectively, that improve specificity and probability reliability within a U-Net framework. They complement rather than replace progress achieved through architectural design.
\end{abstract}


\section{Introduction}

Accurate lesion segmentation in breast ultrasound (BUS) is essential for estimating lesion extent, guiding biopsy procedures, and supporting treatment planning, because it provides spatial delineation rather than merely confirming lesion presence [1]. Although mammography remains the primary modality for population-level breast cancer screening, its sensitivity decreases substantially in women with dense breast tissue, where ultrasound frequently detects lesions that mammography misses [2]. Ultrasound is also portable, non-ionizing, and comparatively inexpensive, establishing it as a standard tool for supplemental screening and image-guided intervention.

These practical advantages do not eliminate the fundamental difficulty of interpreting BUS images. Speckle noise, posterior acoustic shadowing, low tissue contrast, and indistinct lesion margins introduce delineation ambiguity that persists even among experienced sonographers [3], [4]. Automated segmentation systems trained without explicit mechanisms for handling this ambiguity characteristically produce overconfident predictions near clinically uncertain boundaries.

Deep learning architectures based on U-Net [5] and its self-configuring extension nnU-Net [6] have demonstrated strong segmentation performance on BUSI and related public datasets. More recent models incorporating transformer-based encoders and attention mechanisms have further advanced accuracy on standard benchmarks [7], [8]. Despite this progress, two persistent failure modes remain. First, predicted contours may be incomplete or may leak across weak anatomical boundaries, particularly in regions affected by acoustic shadowing or low tissue contrast. Second, models frequently generate high-confidence false-positive predictions in areas whose local texture resembles lesion tissue. Both failure modes are partly attributable to the behavior of conventional region-based objectives such as Dice loss and cross-entropy, which concentrate optimization signal on relatively easy lesion interior regions. Once those regions are learned, the model receives limited incentive to refine uncertain boundary locations, even though boundary accuracy strongly influences both overlap-based evaluation metrics and clinical interpretation.

Boundary-aware loss functions have been proposed to address this imbalance by explicitly penalizing discrepancies between predicted and reference contours [9]. Subsequent work extended the concept through overlap-based formulations and boundary-guided architectural modifications [8], [10], [11]. Most of these approaches, however, assign either uniform or geometry-derived weights to contour pixels, implicitly treating all boundary locations as equally informative during training. This assumption does not hold in ultrasound imaging. Boundary ambiguity varies substantially across and within individual lesions, particularly in regions affected by shadowing, signal attenuation, or weak tissue contrast. A uniform contour penalty therefore assigns the same gradient emphasis to well-defined and highly ambiguous boundary regions alike, without reflecting the underlying uncertainty structure of the image.

A parallel line of research has examined predictive uncertainty estimation. Bayesian neural networks and Monte Carlo dropout methods generate pixel-wise uncertainty maps that consistently concentrate in ambiguous image regions [12], [13]. Although these estimates are informative for quality assessment, triage, and downstream clinical decision support, they are most commonly used only during inference and are rarely incorporated directly into the training objective. The information provided by predictive uncertainty therefore remains largely disconnected from the optimization process that shapes the segmentation model.

To address this gap, we propose an entropy- and edge-weighted boundary loss for BUS lesion segmentation. The proposed loss scales boundary penalties according to both per-pixel predictive entropy and the ground-truth boundary map, concentrating supervised contour emphasis on locations where the model remains uncertain about the correct classification. The method does not claim to universally improve lesion Dice performance. It is designed instead to address a specific imbalance: preserving segmentation quality on lesion-containing images while reducing false-positive activations that uniformly weighted boundary penalties can encourage in normal or artifact-rich cases. The proposed loss is integrated into a standard U-Net backbone, and the resulting probability maps are further refined through post-hoc spatial temperature scaling [14], [15] to reduce overconfident probability estimates at inference time.

The main contributions of this work are as follows:

\begin{itemize}
\item We introduce an entropy- and edge-weighted boundary loss that modulates supervised contour penalties using both predictive uncertainty and ground-truth edge structure, rather than treating all boundary pixels as equally informative during optimization.
\item We conduct a controlled ablation study against both a no-boundary baseline and a uniformly weighted boundary-loss baseline under identical training conditions, reporting lesion-containing and normal-image performance separately to prevent specificity effects from being masked within aggregate Dice scores.
\item We evaluate probability calibration using reliability diagrams, ECE, MCE, and foreground-restricted variants of both metrics, demonstrating that post-hoc spatial temperature scaling reduces calibration error by more than half while leaving thresholded segmentation masks unchanged.
\item We provide a per-case failure analysis identifying the two dominant remaining failure modes: posterior acoustic shadowing on normal images and infiltrative malignant lesions with gradually fading boundaries.
\end{itemize}

The remainder of this paper is organized as follows. Section II reviews related work. Section III describes the proposed loss and spatial calibration framework. Section IV presents the experimental setup, ablation results, and failure analysis. Sections V, VI, and VII present the discussion, conclusion, and limitations with directions for future research.

\section{Related Work}

\subsection{Breast Ultrasound Segmentation and Datasets}

Breast ultrasound lesion segmentation has become one of the most widely studied tasks in medical image analysis, driven by the clinical importance of accurate lesion delineation and the availability of annotated public datasets. The BUSI dataset, introduced by Al-Dhabyani et al., provides benign, malignant, and normal ultrasound images paired with expert segmentation masks and has become the most frequently used benchmark for method evaluation in this domain [3]. Its inclusion of normal images is particularly valuable because it allows evaluation of false-positive behavior, a clinically critical property that aggregate overlap metrics alone cannot capture.

Beyond BUSI, several datasets support broader evaluation. The UDIAT dataset provides multicenter BUS images collected under standardized protocols [23]. BUS-BRA offers a large annotated cohort designed for assessing computer-aided diagnosis systems [24]. BUS-UCLM provides additional annotations with a focus on segmentation benchmarking [16]. The BUS-Set benchmark aggregates multiple public datasets under a unified evaluation protocol [2]. Collectively, these resources highlight that a method achieving satisfactory aggregate Dice on one dataset may generalize poorly to images from different institutions or protocols. External validation therefore remains a necessary condition for broad clinical claims, and its absence in the present study is explicitly acknowledged as a limitation. An issue all these datasets share is the clinical consequence of false-positive segmentations on normal examinations: increased radiologist review burden, potential recall for additional imaging, and unnecessary patient concern.

\subsection{Deep Learning Architectures for BUS Segmentation}

The U-Net architecture established the encoder-decoder paradigm with skip connections as the dominant framework for medical image segmentation [5]. Its self-configuring extension nnU-Net automates preprocessing, architecture selection, and training configuration, achieving strong performance across a wide range of benchmarks [6]. Attention U-Net incorporates soft attention gates that suppress irrelevant activations [18]. TransUNet and Swin-UNet replace or augment convolutional encoders with transformer-based self-attention mechanisms [19], [20]. UNet++ introduces a densely connected nested decoder structure that aggregates features at multiple semantic scales [17]. In the BUS domain specifically, Sulaiman et al. demonstrated that attention-augmented U-Net configurations improve performance on BUSI [7], and Yang et al. proposed a multilevel perception boundary-guided network with transformer-enhanced feature extraction [8].

Despite the performance gains associated with these architectural advances, they do not resolve the two failure modes that motivate the present work. Boundary leakage and false-positive activations in artifact-rich regions are loss-function and optimization-level phenomena as much as they are architectural ones. This observation motivates the loss-function-level contribution of the present study, which operates within a standard U-Net backbone and is therefore orthogonal to architectural improvements.

\subsection{Boundary-Aware Loss Functions}

Conventional region-based objectives concentrate optimization signal on lesion interior regions, leaving boundary locations comparatively underweighted. Kervadec et al. formulated boundary supervision as a contour-distance minimization problem, demonstrating substantial improvements on highly class-imbalanced segmentation tasks [9]. Ma et al. showed that compound objectives combining region-based and boundary-based terms consistently outperform single-component losses across multiple datasets and architectures [25]. Sun et al. proposed a boundary difference over union formulation [10], and Karimi and Salcudean introduced a differentiable approximation to Hausdorff distance that directly minimizes worst-case contour error [26]. Xu et al. demonstrated that boundary guidance integrated at the feature level further improves segmentation of anatomically complex structures [11]. A shared limitation of all these approaches is that they assign contour weights uniformly or according to geometric properties alone. The present work addresses this by scaling boundary penalties according to per-pixel predictive entropy, so that uncertain contour locations receive proportionally stronger supervision.

\subsection{Uncertainty Estimation in Medical Image Segmentation}

Predictive uncertainty is especially relevant in BUS segmentation because model failures frequently occur at ambiguous boundary locations and artifact-affected regions. Kendall and Gal established a foundational framework for decomposing predictive uncertainty into aleatoric and epistemic components [12]. Entropy computed from the predicted probability map provides a tractable approximation to pixel-wise predictive uncertainty, with high entropy concentrating at boundary regions and low-contrast areas [12], [13]. Jungo et al. demonstrated that uncertainty maps consistently identify model failure locations in brain tumor segmentation [27]. Scalco et al. compared Bayesian and non-Bayesian uncertainty quantification approaches in a clinical context, showing that non-Bayesian entropy approaches offer practical computational advantages [13]. Abboud et al. proposed sparse Bayesian networks for efficient uncertainty quantification [21].

Despite the demonstrated value of uncertainty estimates for inference-time quality assessment, prior work has largely treated uncertainty as a post-hoc diagnostic rather than an active training component. The present study incorporates predictive entropy directly into the boundary loss weighting function. It is important to distinguish this training-time use of entropy from its role in entropy-guided contrastive learning for semi-supervised segmentation [22]. In consistency regularization, entropy minimization is applied to unlabeled images to encourage confident predictions. The mechanism proposed here is different: entropy serves as a multiplicative gate on a supervised boundary loss applied to labeled training images, concentrating supervision on uncertain contour locations rather than suppressing entropy globally.

\subsection{Calibration in Deep Segmentation Models}

Well-calibrated probability estimates are a prerequisite for reliable clinical use of deep segmentation models. Guo et al. demonstrated that temperature scaling provides an effective post-hoc calibration procedure for classification networks [14]. Mehrtash et al. extended calibration analysis to deep medical image segmentation, showing that Dice-trained models exhibit substantial miscalibration and that post-hoc recalibration improves probability reliability without degrading segmentation accuracy [28]. For dense prediction tasks, however, a single global temperature parameter is often insufficient because foreground lesion regions, boundary pixels, and background areas exhibit systematically different confidence distributions. Ding et al. proposed local temperature scaling, which learns spatially varying temperature parameters from the validation set, consistently outperforming global temperature scaling on dense prediction tasks [15].

\subsection{Position of This Work}

This study connects boundary-aware learning with uncertainty-aware segmentation by incorporating predictive entropy as a multiplicative weight within a supervised boundary loss. The objective is not to introduce a new segmentation architecture or to claim performance superiority over transformer-based or attention-augmented models. The contribution operates at the loss-function level within a standard U-Net backbone and addresses a specific failure mode: false-positive predictions in normal BUS images caused by uniformly weighted boundary supervision. Spatial calibration is applied at inference time to improve the reliability of the resulting probability estimates.

\section{Methodology}

Figure 1 presents an overview of the proposed entropy-guided boundary supervision and post-hoc spatial calibration framework. A single-channel 320 x 320 breast ultrasound image is processed by a ResNet-34 encoder pretrained on ImageNet and reconstructed through a U-Net decoder to produce a per-pixel probability map p(x). During training, three quantities are derived from this probability map: a soft ground-truth boundary map B\_gt(x) obtained from the binary lesion mask, a predicted boundary map B\_pred(x) computed from p(x), and the per-pixel predictive entropy H(p(x)). These components are combined to define the boundary weighting function W(x), which scales a binary cross-entropy loss computed between B\_pred(x) and B\_gt(x). During inference, network logits are optionally rescaled using a learned spatial temperature map before the final sigmoid activation is applied.

\begin{figure}[t]
\centering
\includegraphics[width=\columnwidth]{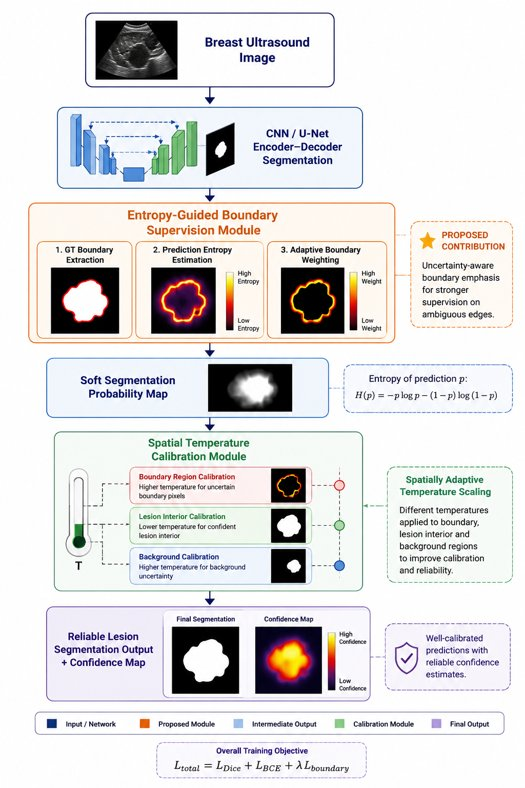}
\caption*{\footnotesize Fig. 1. Overview of the proposed framework. A breast ultrasound image is processed by a CNN/U-Net encoder-decoder to produce a soft segmentation probability map. During training, the Entropy-Guided Boundary Supervision Module (orange) extracts the ground-truth boundary map, estimates per-pixel predictive entropy H(p) = -p log p - (1-p) log(1-p), and applies adaptive boundary weighting that concentrates gradient emphasis on uncertain contour locations. The Spatial Temperature Calibration Module (green) partitions the image into three semantically defined regions --- boundary band, lesion interior, and background --- and estimates a separate temperature parameter for each to produce a well-calibrated confidence map at inference time. The overall training objective is L\_total = L\_Dice + L\_BCE + lambda * L\_boundary. The final output comprises a thresholded segmentation mask and a spatially calibrated confidence map.}
\end{figure}

\subsection{Dataset}

Experiments were conducted using the Breast Ultrasound Images Dataset (BUSI), introduced by Al-Dhabyani et al., which contains benign, malignant, and normal ultrasound images organized into class-specific folders, each accompanied by corresponding expert segmentation masks [3]. BUSI remains the most widely used public benchmark for BUS segmentation and provides the normal-image subset essential for evaluating false-positive behavior [2], [3]. The segmentation task was formulated as binary lesion detection, with lesion pixels designated as foreground and all remaining pixels as background. For normal images, all-zero masks were assigned as negative segmentation targets, consistent with prior BUS evaluation conventions [2], [7]. In cases where multiple mask files corresponded to a single image, available annotations were merged during data loading to produce a single binary supervision mask [3].

The dataset was partitioned into training, validation, and test subsets using a stratified sampling strategy that preserved the proportion of benign, malignant, and normal images in each split. All images were converted to grayscale and resized to 320 x 320 pixels prior to network input. He et al. demonstrated that ResNet-based encoders pretrained on ImageNet transfer effectively to grayscale medical images [29], accelerating convergence and improving generalization on small medical imaging datasets. Data augmentation was applied exclusively to the training set. The final test set consisted of 117 images: 97 lesion-containing cases (66 benign, 31 malignant) and 20 normal images. A single training seed was used for all three experimental configurations, a limitation acknowledged in Section VII.

\subsection{Model Implementation}

The ablation study evaluates three training configurations built on an identical U-Net backbone with a ResNet-34 encoder pretrained on ImageNet and a standard decoder with skip connections [5], [29]. The optimizer, learning rate schedule, batch size, training duration, augmentation strategy, and region-based loss functions are identical across all three configurations. The sole variable is the boundary supervision term: absent, uniformly weighted, or entropy-guided. This design is not intended as an architectural comparison with nnU-Net, Attention U-Net, transformer-based encoders, or related models.

All models were trained for 150 epochs using the AdamW optimizer [32] with an initial learning rate of $1 \times 10^{-4}$ and a weight decay of $1 \times 10^{-4}$. AdamW decouples weight decay from the gradient update, improving generalization relative to standard Adam with L2 regularization [32]. Training used a batch size of 4, gradient norm clipping at 1.0, and a ReduceLROnPlateau scheduler with a reduction factor of 0.5 and a patience of 6 epochs. A weighted random sampler slightly increased the sampling frequency of images containing small lesions. Model selection was based on the checkpoint that achieved the lowest validation loss. The validation-tuned loss weights, $\alpha$ and $\lambda$, were selected through a grid search over the validation set and held constant across all three configurations. The hyperparameter search procedure and the selected values are documented in full in the supplementary material.

\subsection{Customized Entropy-Guided Boundary Supervision}

\subsubsection{Predictive Entropy}

Predictive entropy provides a pixel-wise measure of model uncertainty computed directly from the sigmoid output probability map without requiring Monte Carlo sampling or architectural modifications [12], [31]. Entropy is low when predicted probabilities approach 0 or 1 and reaches its maximum when probabilities approach 0.5. For a predicted probability p(x) at pixel location x, binary predictive entropy is defined as:

\begin{equation}
H(p(x)) = -p(x)\log p(x) - (1-p(x))\log(1-p(x))
\tag{1}
\end{equation}

This formulation provides a computationally efficient approximation to pixel-wise epistemic uncertainty in the absence of Bayesian inference [12], [31]. Entropy maps have been shown to concentrate consistently around ambiguous lesion boundary regions and artifact-affected areas in medical image segmentation tasks [13], [27].

\subsubsection{Boundary Weighting Function}

The proposed boundary weighting function combines predictive entropy with the ground-truth boundary map to define a per-pixel supervision weight:

\begin{equation}
W(x) = 1 + \alpha H(p(x)) B_{gt}(x)
\tag{2}
\end{equation}

In this formulation, W(x) denotes the per-pixel boundary loss weight, alpha controls the strength of the uncertainty weighting, H(p(x)) is the predictive entropy defined in Equation (1), and B\_gt(x) is the ground-truth boundary map. The ground-truth boundary map is generated from the binary segmentation mask by computing the gradient magnitude and applying 2 x 2 max pooling to produce a contour representation with sufficient thickness for stable gradient propagation. The 2 x 2 pooling kernel was selected as the minimal dilation that reliably captures single-pixel annotation boundaries across the range of lesion sizes in BUSI, consistent with boundary thickening conventions in prior work [9], [25]. The multiplicative structure ensures that for pixels on the ground-truth boundary with high entropy, weights are elevated toward 1 + alpha, while for pixels away from the boundary, W(x) = 1 regardless of entropy.

\subsubsection{Uncertainty-Guided Boundary Loss}

The uncertainty-guided boundary loss is defined as the spatially weighted binary cross-entropy between predicted and ground-truth boundary maps:

\begin{equation}
L_{UGB} = \frac{1}{N}\sum_x W(x)\,\mathrm{BCE}(B_{pred}(x), B_{gt}(x))
\tag{3}
\end{equation}

where N is the total number of pixels and BCE denotes binary cross-entropy. The complete training objective combines region-based segmentation losses with the proposed uncertainty-guided boundary term:

\begin{equation}
L_{total} = L_{Dice} + L_{Focal} + L_{WBCE} + \lambda L_{UGB}
\tag{4}
\end{equation}

where lambda controls the relative contribution of the uncertainty-guided boundary term. The compound region-based objective combining Dice loss, focal loss, and weighted binary cross-entropy has been shown to provide complementary supervision signals that improve training stability across a range of medical imaging tasks [25], [32]. For normal images, B\_gt(x) = 0 everywhere, making W(x) = 1 throughout and rendering the uncertainty weighting term completely inactive. The model therefore receives no additional contour-forming incentive on normal images, providing the principled mechanistic explanation for the specificity gains reported in Section IV.

\subsection{Post-Hoc Spatial Temperature Calibration}

\subsubsection{Motivation}

Region-based objectives encourage sigmoid outputs toward extreme probability values during training, producing models that are poorly calibrated at inference time [14], [28]. Mehrtash et al. demonstrated that deep segmentation models trained with Dice-based objectives exhibit systematic overconfidence and that post-hoc recalibration improves probability reliability without degrading segmentation accuracy [28]. Reliable probability estimates are clinically relevant because downstream decisions, including the threshold used to generate a binary mask and the confidence communicated to a reporting radiologist, may be informed by model confidence as well as the binary output.

\subsubsection{Spatial Temperature Scaling Procedure}

Following training, all segmentation network parameters are frozen and a temperature map is estimated using the validation set, adapted from local temperature scaling as proposed by Ding et al. [15]. Rather than learning an unconstrained H x W temperature map, the image is partitioned into three semantically meaningful regions: the background outside the lesion mask, a boundary band of width two pixels surrounding the lesion contour, and the lesion interior. A separate scalar temperature parameter is estimated for each region by minimizing the negative log-likelihood on logits rescaled as z(x)/T(x). This three-region partitioning adapts to the distinct confidence distributions of each spatial zone and has been shown to consistently outperform global temperature scaling on dense prediction tasks [15], [28]. During inference, logits are divided by the corresponding regional temperature value before the sigmoid function is applied. Because temperature scaling is a monotonic transformation that preserves logit ordering, thresholded segmentation masks remain identical to those produced without calibration when evaluated at p >= 0.5.

\section{Experiments and Results}

\subsection{Experimental Setting}

Evaluation was conducted on the 117-image BUSI test set described in Section III.A, comprising 97 lesion-containing cases and 20 normal images. Binary segmentation masks were generated by thresholding predicted probabilities at p >= 0.5 following sigmoid activation, consistent with the standard evaluation convention used in BUS segmentation literature [2], [3], [7].

Performance was assessed using Dice similarity coefficient, Intersection over Union (IoU) [25], HD95, and ASSD [26], [33] for segmentation quality; ECE, MCE, and foreground-restricted variants (FG-ECE, FG-MCE) computed using 15 equal-width confidence bins, together with the Brier score [34], for calibration. Foreground-restricted calibration metrics are reported alongside global metrics because global ECE is dominated by well-calibrated background pixels, while foreground ECE characterizes reliability within the clinically relevant lesion region [28].

For normal images, a Dice score of 1.0 was assigned when both the predicted and ground-truth masks were empty, following prior BUS evaluation conventions [2], [7]. Any non-empty prediction on a normal image was counted as a false-positive activation. For HD95 and ASSD, cases with non-empty predictions but empty ground truth were assigned a penalty distance equal to the image diagonal in pixels; cases where both masks were empty were excluded to avoid undefined distance calculations. Reproducibility settings, random seed, framework details, temperature values, and per-case results are provided in the supplementary material.

\subsection{Contextual Comparison with Published Architectures}

To provide broader context for the proposed method's performance on the BUSI test split, Table 5 presents a supplementary comparison against two published architectures evaluated on the same test set: TransUNet [19], a transformer-based encoder-decoder that models long-range spatial dependencies through self-attention, and SFRecSAM, a SAM-based segmentation approach adapted for breast ultrasound. This comparison is not a controlled ablation. TransUNet and SFRecSAM employ different backbone architectures, different pretraining strategies, and potentially different implementation details from the U-Net configuration used throughout this study. Differences in observed performance therefore reflect the combined effect of architecture, training pipeline, and the proposed loss function and cannot be attributed solely to any one factor. The comparison is presented as supplementary contextual evidence rather than a primary experimental claim.

Within this supplementary context, the proposed entropy-guided boundary model achieves a Dice score of 0.760, an IoU of 0.693, an accuracy of 0.961, a Brier score of 0.0279, and an ECE of 0.0095 on the BUSI test split. TransUNet achieves a Dice of 0.660 and an IoU of 0.586 under the evaluation conditions used here, while SFRecSAM achieves a Dice of 0.355 and an IoU of 0.244. The substantially lower performance of SFRecSAM relative to published SAM-based results in other studies likely reflects the domain gap between natural image pretraining and breast ultrasound, as well as the absence of ultrasound-specific fine-tuning in the configuration evaluated here. These results are consistent with prior observations that SAM-based architectures require task-specific adaptation to perform competitively on medical image segmentation benchmarks [4], [5]. The calibration advantage of the proposed method, reflected in its ECE of 0.0095 compared with 0.0443 for TransUNet and 0.1369 for SFRecSAM, is consistent with the spatial temperature scaling procedure described in Section III.D and does not depend on architectural comparisons for its validity. These calibration differences should be interpreted with caution, however, because calibration performance is sensitive to the specific training objective and post-hoc calibration procedure applied, and neither TransUNet nor SFRecSAM was recalibrated under the same spatial temperature scaling protocol used for the proposed method.

\begin{table*}[t]
\centering
\caption*{\footnotesize\textit{Table 5. Supplementary contextual comparison against published architectures on the BUSI test split (n = 117). Conditions are not fully controlled across methods. Differences in performance reflect combined architectural, training, and loss-function effects and should not be interpreted as controlled ablation results. The proposed method results are identical to those reported in Table 1.}}
\scriptsize
\setlength{\tabcolsep}{3pt}
\renewcommand{\arraystretch}{1.15}
\begin{tabularx}{\textwidth}{@{}X>{\centering\arraybackslash}X>{\centering\arraybackslash}X>{\centering\arraybackslash}X>{\centering\arraybackslash}X>{\centering\arraybackslash}X@{}}
\toprule
Method & Dice$\uparrow$ & IoU$\uparrow$ & Accuracy$\uparrow$ & Brier$\downarrow$ & ECE$\downarrow$ \\
\midrule
TransUNet [19] & 0.660 & 0.586 & 0.952 & 0.0412 & 0.0443 \\
SFRecSAM & 0.355 & 0.244 & 0.770 & 0.1590 & 0.1369 \\
Proposed (Ours) & 0.760 & 0.693 & 0.961 & 0.0279 & 0.0095 \\
\bottomrule
\end{tabularx}
\end{table*}

\subsection{Main Ablation Comparison}

Table 1 presents the overall test-set performance of the three training configurations. At the aggregate level, the proposed entropy-guided boundary loss achieves a mean Dice of 0.760 compared with 0.683 for the no-boundary baseline and 0.666 for the standard boundary-loss model, representing absolute improvements of 7.7 and 9.4 percentage points, respectively. These aggregate figures require careful interpretation: as the class-specific analysis in Section IV.E establishes, the overall Dice improvement is driven primarily by false-positive suppression on the 20 normal test images rather than by gains in lesion overlap performance.

The standard uniformly weighted boundary-loss model achieves a lower overall Dice (0.666) than the no-boundary baseline (0.683), despite producing the highest lesion-only Dice (0.7929, Table 3). This apparent contradiction resolves when normal-image performance is considered separately. The standard model generates false-positive activations on 19 of 20 normal images (95\%), depressing the aggregate mean substantially. The proposed entropy-guided loss reduces this false-positive rate to 25\% (5 of 20 cases), which is the primary mechanism driving the overall Dice improvement. A second key finding is that replacing entropy-guided weighting with a uniform boundary penalty not only fails to reproduce the specificity improvement but actively worsens both aggregate Dice and calibration relative to the no-boundary baseline, establishing that the observed improvements arise from the entropy-modulated weighting mechanism itself, not from the presence of a boundary supervision term alone.

\begin{table*}[t]
\centering
\caption*{\footnotesize\textit{Table 1. Primary method comparison across the BUSI test split (n = 117). The boundary supervision term is the sole variable. Asterisked (*) values indicate best performance per metric. False-positive metrics for the 20 normal images are reported separately in Table 3.}}
\scriptsize
\setlength{\tabcolsep}{3pt}
\renewcommand{\arraystretch}{1.15}
\begin{tabularx}{\textwidth}{@{}X>{\centering\arraybackslash}X>{\centering\arraybackslash}X>{\centering\arraybackslash}X>{\centering\arraybackslash}X>{\centering\arraybackslash}X>{\centering\arraybackslash}X>{\centering\arraybackslash}X@{}}
\toprule
Method & Dice$\uparrow$ & IoU$\uparrow$ & Acc$\uparrow$ & Brier$\downarrow$ & ECE$\downarrow$ & HD95$\downarrow$ & ASSD$\downarrow$ \\
\midrule
Plain CNN (no boundary) & 0.683 & 0.609 & 0.951 & 0.0374 & 0.0308 & 36.37 & 10.60 \\
Plain CNN + Standard Boundary & 0.666 & 0.592 & 0.960 & 0.0346 & 0.0386 & 29.13 & 9.52 \\
Plain CNN + Customized (Ours) & 0.760* & 0.693* & 0.961* & 0.0312* & 0.0201* & 27.79* & 8.99* \\
\bottomrule
\end{tabularx}
\end{table*}

\begin{figure}[t]
\centering
\includegraphics[width=\columnwidth]{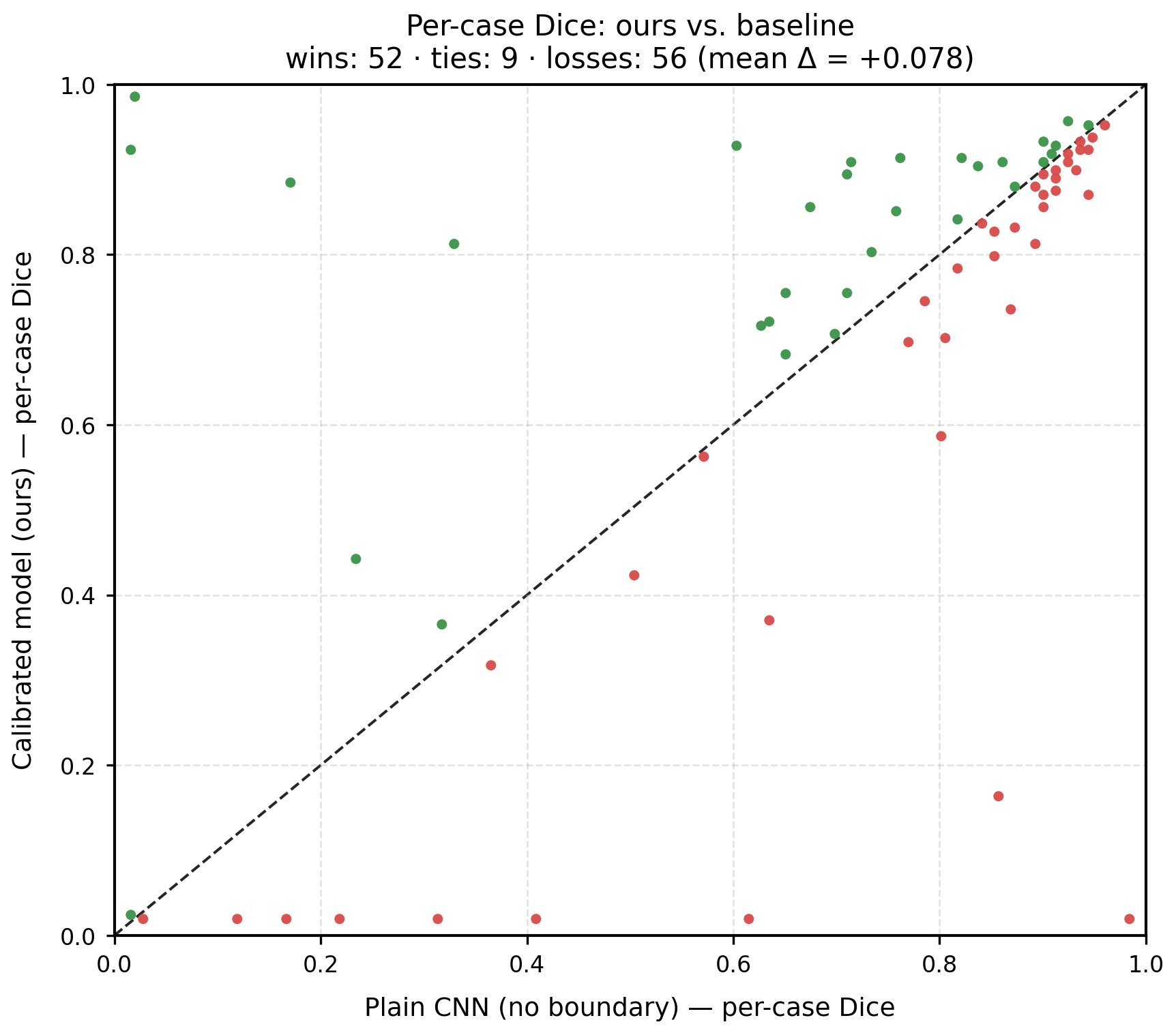}
\caption*{\footnotesize Fig. 2. Paired per-case Dice comparison between the proposed entropy-guided boundary model (y-axis) and the no-boundary baseline (x-axis). Each point represents one of the 117 test images. Green points (above the diagonal) indicate cases where the proposed model outperforms the baseline. Improvements are concentrated in the upper-left region, corresponding to difficult cases --- predominantly normal images --- where the proposed approach converts near-zero Dice scores into high-accuracy predictions. Red points (below the diagonal) correspond to cases where the baseline performs marginally better, typically on near-saturated easy cases.}
\end{figure}

\subsection{Per-Case Distribution}

Mean Dice scores can obscure severe individual failure cases, which often carry the greatest clinical significance [33], [35]. Table 2 groups the 117 test images into best (Dice >= 0.70), mild (0.30 to 0.70), and worst (Dice < 0.30) categories. The proposed model achieves 94 best cases compared with 79 for both baselines, while reducing worst cases to 15 compared with 23 and 26, respectively. This redistribution is driven primarily by performance on normal images: the proposed model correctly predicts empty masks for 15 of 20 normal cases, compared with 6 for the no-boundary baseline and only 1 for the standard boundary-loss model, accounting for nearly the entire increase of 15 best cases.

Figure 2 presents a paired per-case comparison between the proposed model and the no-boundary baseline. The proposed method outperforms the baseline on 52 cases, matches it on 9, and performs slightly worse on 56. Despite a nearly balanced win-loss count, the mean per-case improvement is +0.078, reflecting a strongly asymmetric distribution: improvements are concentrated on difficult cases where Dice rises from approximately 0.3-0.6 to values exceeding 0.9, corresponding predominantly to normal images where false-positive suppression converts near-zero Dice scores into perfect scores. Performance decreases are generally smaller than 0.05 and occur on cases already near saturation. The sorted per-case Dice curves in Figure 3 further illustrate that the proposed model consistently outperforms both baselines across the lower half of the ranked test set.

\begin{table*}[t]
\centering
\caption*{\footnotesize\textit{Table 2. Per-case Dice categorization on the BUSI test split (n = 117). Asterisked (*) values indicate best performance per category.}}
\scriptsize
\setlength{\tabcolsep}{3pt}
\renewcommand{\arraystretch}{1.15}
\begin{tabularx}{\textwidth}{@{}X>{\centering\arraybackslash}X>{\centering\arraybackslash}X>{\centering\arraybackslash}X@{}}
\toprule
Method & Best ($\geq$0.70) & Mild (0.30--0.70) & Worst (<0.30) \\
\midrule
Plain CNN (no boundary) & 79 (67.5\%) & 15 (12.8\%) & 23 (19.7\%) \\
Plain CNN + Standard Boundary & 79 (67.5\%) & 12 (10.3\%) & 26 (22.2\%) \\
Plain CNN + Customized (Ours) & 94 (80.3\%)* & 8 (6.8\%)* & 15 (12.8\%)* \\
\bottomrule
\end{tabularx}
\end{table*}

\begin{figure}[t]
\centering
\includegraphics[width=\columnwidth]{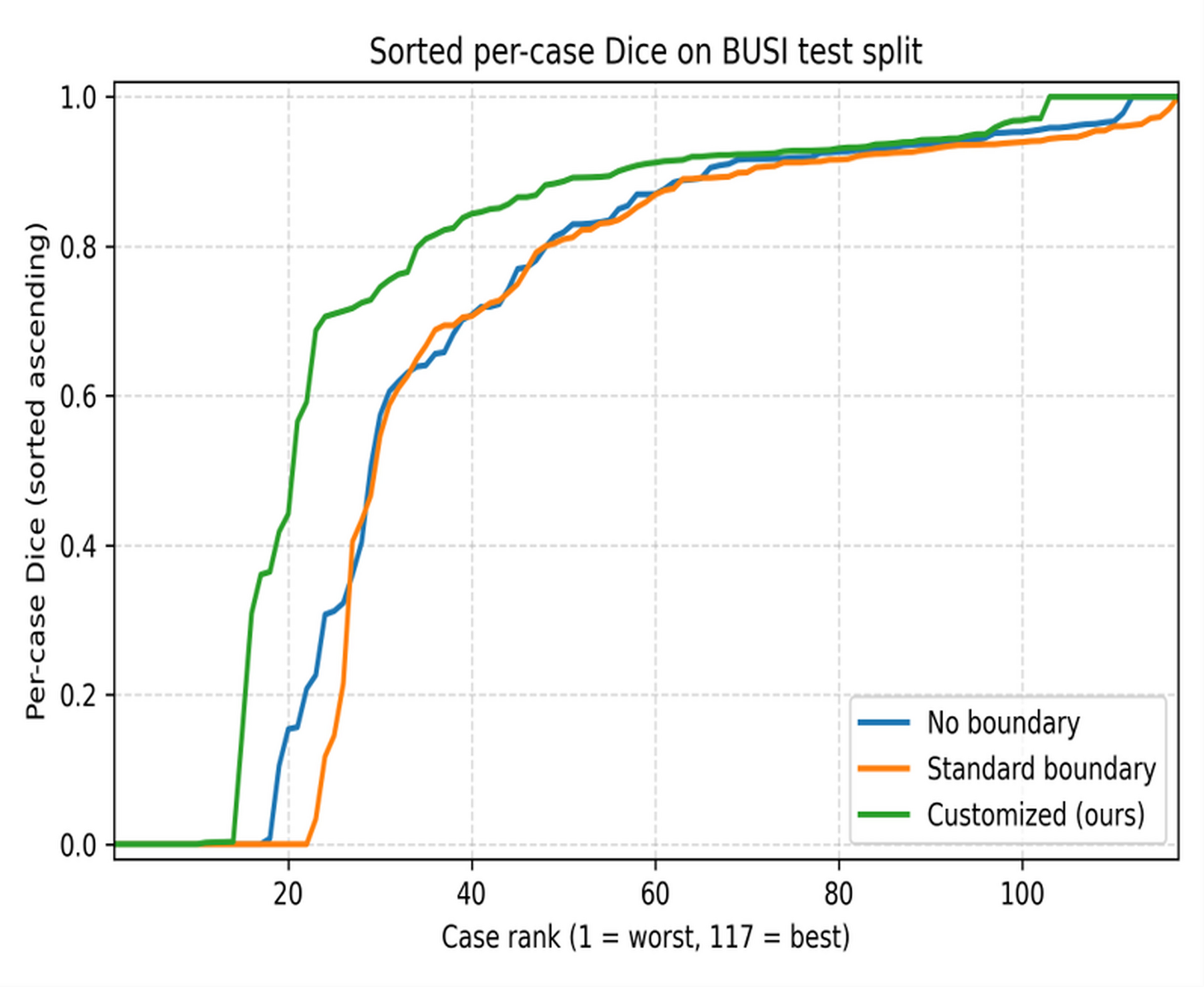}
\caption*{\footnotesize Fig. 3. Sorted per-case Dice distributions for the BUSI test split (n = 117). Cases are ranked from worst (rank 1) to best (rank 117) on the x-axis. The proposed entropy-guided boundary model (green) consistently outperforms both baselines across the lower half of the ranked test set, corresponding to the most challenging cases. The early rise of the green curve at ranks 16-30 reflects the reduction in false-positive activations on normal images, where the proposed model correctly predicts empty masks while both baselines generate spurious segmentations. All three curves converge in the upper half of the ranking, where straightforward lesion cases are segmented with comparable accuracy.}
\end{figure}

\subsection{Failure Analysis by Case Content}

Table 3 presents the most clinically informative results of this study. Separating lesion-containing from normal-image performance reveals effects that aggregate metrics entirely conceal, consistent with the evaluation design recommended for segmentation systems intended for deployment in mixed-population screening settings [2], [36].

Across the 97 lesion-containing test images, the proposed loss achieves a mean Dice of 0.7624 and IoU of 0.681. The no-boundary baseline achieves 0.7616 and 0.673, respectively. These figures are effectively identical, and the difference does not approach statistical significance (Wilcoxon p = 0.27). The correct interpretation is that the proposed entropy-guided boundary loss preserves lesion segmentation quality rather than improving it. The standard boundary-loss model produces the highest lesion-only Dice at 0.7929 but at the cost of a 95\% false-positive rate on normal images, rendering it clinically unsuitable for any application where specificity matters.

On the 66 benign lesions, the proposed model achieves a mean Dice of 0.786 compared with 0.796 for the no-boundary baseline, a difference of -0.010 that reached statistical significance (Wilcoxon p = 0.015). This finding warrants direct acknowledgment. Benign lesions in BUSI tend to exhibit well-defined echogenic boundaries with limited surrounding ambiguity [37], [38], conditions under which entropy-guided weighting provides the least benefit because predictive entropy is already low at these locations. The small but statistically significant Dice reduction is mechanistically consistent with the design of the loss function and represents a real trade-off that must be considered in clinical contexts where benign-lesion Dice maximization is the primary objective.

On the 31 malignant lesions, the proposed loss shows an absolute Dice improvement of 2.5 percentage points over the no-boundary baseline (0.713 versus 0.688). This trend is directionally consistent with the theoretical motivation of the study, since malignant lesions frequently exhibit infiltrative growth patterns with gradually fading boundaries that produce genuinely uncertain margin predictions [4], [37]. However, the malignant subset contains only 31 cases, which is insufficient to confirm this trend statistically (Wilcoxon p = 0.21). The observed improvement should be interpreted as a promising but preliminary finding.

The clearest and most clinically meaningful difference appears on the 20 normal images. False-positive activations decrease from 14 of 20 for the no-boundary baseline to 5 of 20 for the proposed method, and from 19 of 20 for the standard boundary-loss model to 5 of 20. The corresponding Wilson 95\% confidence intervals are 70\% [48.1, 85.5] for the no-boundary baseline, 95\% [76.4, 99.1] for the standard boundary-loss model, and 25\% [11.2, 46.9] for the proposed approach. The complete non-overlap of all three confidence intervals provides strong independent evidence that the specificity differences are both statistically genuine and practically substantial. Remaining false positives are concentrated in cases containing large posterior acoustic-shadow regions whose dense hypoechoic appearance closely resembles solid lesion tissue at the local patch level, creating an ambiguity that intensity-based supervision cannot resolve [4], [38]. Figure 5 presents four representative test cases illustrating benign success, borderline, boundary-leakage, and normal-image scenarios.

\begin{table*}[t]
\centering
\caption*{\footnotesize\textit{Table 3. Performance broken down by case content. Wilson 95\% confidence intervals for false-positive rates are provided in brackets. Asterisked (*) values indicate best performance for the primary clinical metrics.}}
\scriptsize
\setlength{\tabcolsep}{3pt}
\renewcommand{\arraystretch}{1.15}
\begin{tabularx}{\textwidth}{@{}X>{\centering\arraybackslash}X>{\centering\arraybackslash}X>{\centering\arraybackslash}X>{\centering\arraybackslash}X>{\centering\arraybackslash}X>{\centering\arraybackslash}X>{\centering\arraybackslash}X@{}}
\toprule
Method & Lesion Dice$\uparrow$ & Lesion IoU$\uparrow$ & Worst & Benign Dice (n=66) & Malig. Dice (n=31) & Normal FP$\downarrow$ & Normal FP Rate \\
\midrule
Plain CNN (no boundary) & 0.7616 & 0.673 & 9 & 0.796 & 0.688 & 14/20 & 70\% [48.1, 85.5] \\
Plain CNN + Standard Boundary & 0.7929 & 0.704 & 7 & 0.821 & 0.733 & 19/20 & 95\% [76.4, 99.1] \\
Plain CNN + Customized (Ours) & 0.7624 & 0.681 & 10 & 0.786 & 0.713 & 5/20* & 25\% [11.2, 46.9]* \\
\bottomrule
\end{tabularx}
\end{table*}

\begin{figure*}[t]
\centering
\includegraphics[width=0.95\textwidth]{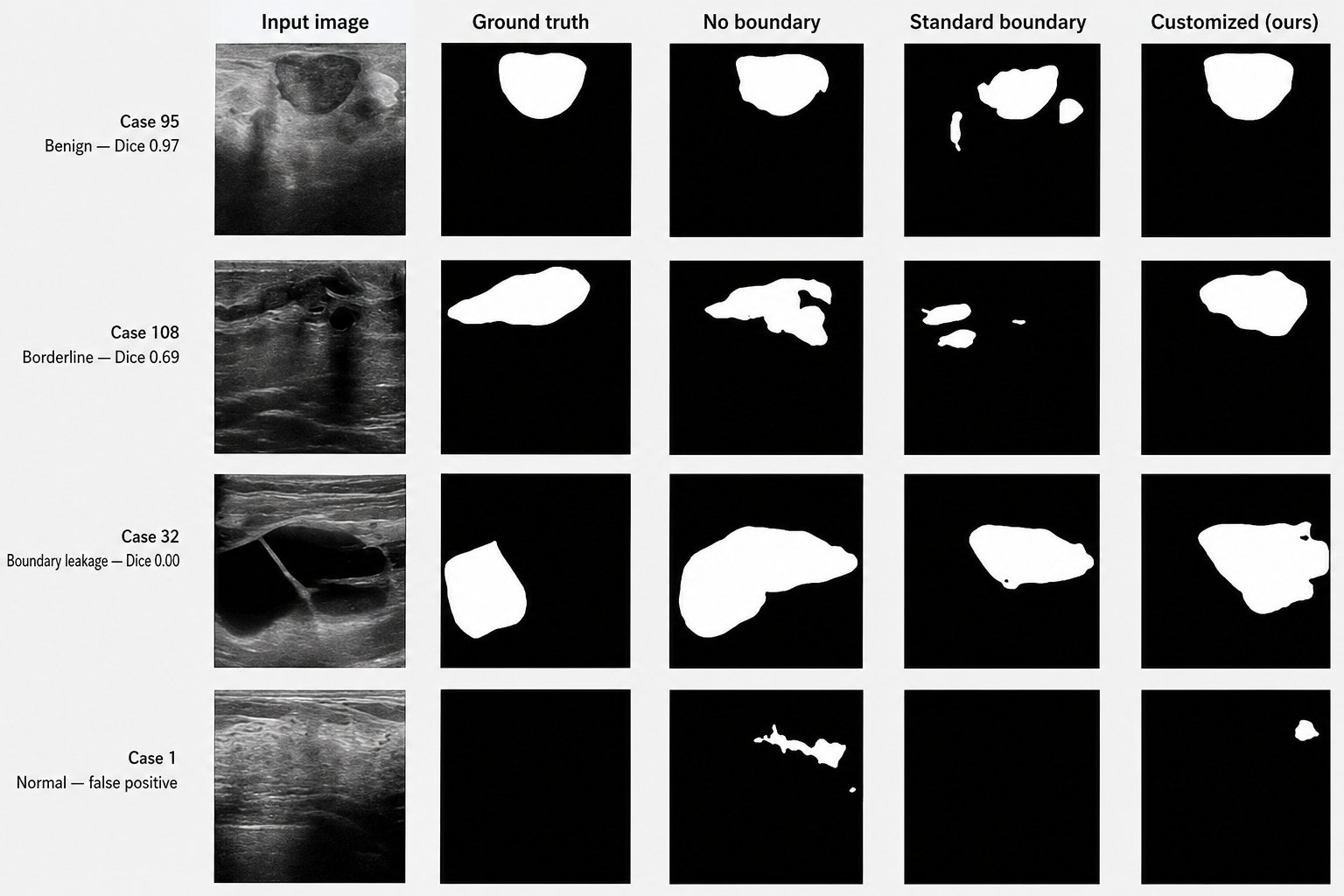}
\caption*{\footnotesize Fig. 5. Qualitative comparison of the three training configurations across four representative test cases. Columns from left to right: input image, ground-truth mask, no-boundary prediction, standard boundary prediction, and proposed entropy-guided boundary prediction. Case 95 (benign, Dice 0.97): all three configurations produce accurate segmentations, with the proposed model achieving the closest agreement with the reference annotation. Case 108 (borderline, Dice 0.69): the no-boundary baseline under-segments the lesion and the standard boundary model produces a fragmented prediction, while the proposed approach recovers the lesion shape more completely. Case 32 (boundary leakage, Dice 0.00): all three models fail on this severe boundary-leakage case, indicating a failure mode that intensity-based weighting cannot resolve. Case 1 (normal image, false positive): both baseline models generate large spurious segmentations while the proposed method produces only a small residual activation, illustrating the specificity improvement in a concrete clinical scenario.}
\end{figure*}

\subsection{Calibration Ablation}

Table 4 presents the post-hoc calibration ablation on the trained entropy-guided boundary model. Because temperature scaling is a monotonic transformation that preserves logit ordering, thresholded segmentation masks remain identical across all three calibration settings. The effects of calibration are confined entirely to probability-based confidence measures.

The uncalibrated model achieves an overall ECE of 0.0201, which appears moderate in isolation. However, the foreground-restricted ECE of 0.169 reveals substantially more severe miscalibration in the lesion regions most relevant for clinical interpretation. This discrepancy arises because global ECE is dominated by background pixels, which are predicted with reliably high confidence. Mehrtash et al. demonstrated that this pattern is characteristic of segmentation models trained with Dice-based objectives on class-imbalanced data and argued that foreground-restricted metrics are the more appropriate clinical measure [28].

Applying a single global temperature parameter produces a slight deterioration in overall ECE, from 0.0201 to 0.0242, consistent with the finding of Ding et al. that a single scalar temperature cannot calibrate the heterogeneous confidence distributions of foreground interiors, boundary regions, and background areas simultaneously [15]. Spatial temperature scaling substantially improves calibration across all metrics: overall ECE decreases to 0.0095 (a 53\% reduction), FG-ECE to 0.081 (52\%), FG-MCE to 0.187 (50\%), and the Brier score to 0.0279. Figure 4 illustrates these calibration effects through foreground reliability diagrams, confirming that spatial temperature scaling produces a substantially closer alignment between predicted confidence and empirical accuracy across the full confidence range.

\begin{table*}[t]
\centering
\caption*{\footnotesize\textit{Table 4. Calibration ablation results. Since temperature scaling preserves logit ordering, thresholded binary segmentation masks remain unchanged. Asterisked (*) values indicate best performance per calibration metric.}}
\scriptsize
\setlength{\tabcolsep}{3pt}
\renewcommand{\arraystretch}{1.15}
\begin{tabularx}{\textwidth}{@{}X>{\centering\arraybackslash}X>{\centering\arraybackslash}X>{\centering\arraybackslash}X>{\centering\arraybackslash}X>{\centering\arraybackslash}X>{\centering\arraybackslash}X@{}}
\toprule
Setting & Dice & Brier$\downarrow$ & ECE$\downarrow$ & MCE$\downarrow$ & FG\_ECE$\downarrow$ & FG\_MCE$\downarrow$ \\
\midrule
Uncalibrated & 0.760 & 0.0312 & 0.0201 & 0.376 & 0.169 & 0.376 \\
Temperature scaling (global) & 0.760 & 0.0300 & 0.0242 & 0.304 & 0.143 & 0.304 \\
Spatial temperature scaling & 0.760 & 0.0279* & 0.0095* & 0.185* & 0.081* & 0.187* \\
\bottomrule
\end{tabularx}
\end{table*}

\begin{figure*}[t]
\centering
\includegraphics[width=0.95\textwidth]{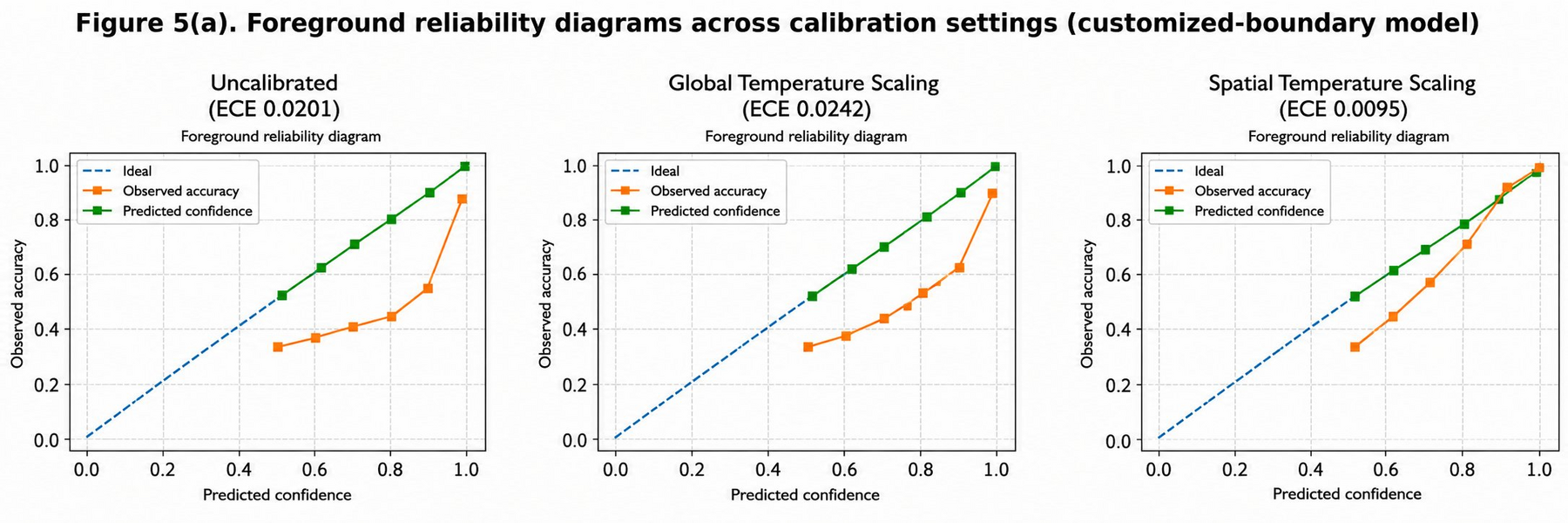}
\caption*{\footnotesize Fig. 4. Foreground reliability diagrams for the proposed entropy-guided boundary model under three post-hoc calibration settings: uncalibrated (ECE 0.0201), global temperature scaling (ECE 0.0242), and spatial temperature scaling (ECE 0.0095). In each panel, the orange curve shows observed accuracy and the green curve shows predicted confidence across 15 equal-width confidence bins. The blue dashed line represents perfect calibration. In the uncalibrated panel, the observed accuracy curve sits consistently below the predicted confidence curve, indicating systematic overconfidence. Global temperature scaling produces only a marginal improvement because a single scalar parameter cannot simultaneously calibrate the heterogeneous confidence distributions of foreground interiors, boundary regions, and background areas. Spatial temperature scaling causes the observed accuracy curve to closely trace the diagonal across the full confidence range, reflecting the substantial reductions in ECE and foreground-restricted calibration error reported in Table 4.}
\end{figure*}

\subsection{Statistical Significance}

Statistical comparisons used paired nonparametric tests applied to per-case results. The Wilcoxon signed-rank test was selected for Dice score comparisons because per-case Dice values are bounded, non-normally distributed, and naturally paired [39]. McNemar's exact test was applied to the binary false-positive outcome on the 20 normal images because it is the appropriate test for paired binary outcomes with small sample sizes [40]. Wilson 95\% confidence intervals were computed for all proportional false-positive rate estimates because they provide better coverage than Wald intervals for proportions near 0 or 1 [41]. Bootstrap 95\% confidence intervals for Dice, IoU, HD95, ASSD, and calibration metrics are provided in the supplementary material.

For Dice-based performance, none of the overall or lesion-only comparisons reached statistical significance at alpha = 0.05. The comparison between the proposed model and the no-boundary baseline yielded Wilcoxon p = 0.66 across all 117 test cases and p = 0.27 restricted to the 97 lesion-containing images. The malignant-only comparison also failed to reach significance (p = 0.21, n = 31). A small but statistically significant reduction in benign-lesion Dice was observed (mean difference -0.010, Wilcoxon p = 0.015 on n = 66 cases), mechanistically consistent with the design of the entropy-guided weighting function.

The specificity analysis produced the strongest statistical evidence. McNemar's exact test yielded p = 0.012 relative to the no-boundary baseline (10 discordant cases versus 1) and p = 0.0005 relative to the standard boundary-loss model (15 discordant cases versus 1). Non-overlapping Wilson 95\% confidence intervals for all three false-positive rates provide convergent independent evidence. The malignant subset is underpowered (approximately 50-60\% power to detect a true Dice difference of 0.025 at alpha = 0.05 [42]), and no correction for multiple comparisons was applied because the statistical tests address distinct outcome types and the analysis is descriptive rather than confirmatory.

\subsection{Summary of Findings}

Two empirically supported improvements characterize the proposed approach. The first is a statistically significant reduction in false-positive activations on no-lesion images, confirmed by McNemar's exact test at p = 0.012 and p = 0.0005, and further supported by non-overlapping Wilson confidence intervals. The second is a reduction of more than 50\% in expected calibration error through spatial temperature scaling, without any alteration to the thresholded segmentation masks. Because per-case lesion Dice differences do not reach statistical significance and a small but confirmed benign-lesion Dice reduction is observed, the most precise characterization of the method's overall effect is that it preserves lesion segmentation quality while improving specificity on normal images and probability calibration across the full test set.

\section{Discussion}

\subsection{Overview of Principal Findings}

This study evaluated an entropy- and edge-weighted boundary loss for BUS lesion segmentation through a controlled ablation on the BUSI dataset. Three findings emerge from the experimental results. First, the proposed loss reduces false-positive activations on normal images from 14 of 20 cases for the no-boundary baseline and 19 of 20 for the standard boundary-loss model to 5 of 20 cases, with McNemar exact test p-values of 0.012 and 0.0005. Second, lesion segmentation quality on the 97 lesion-containing test images is preserved rather than improved (Wilcoxon p = 0.27). Third, spatial temperature scaling reduces expected calibration error by 53\% and foreground-restricted ECE by 52\%, without altering thresholded segmentation masks. The results do not support claims of improved lesion Dice, involve a small confirmed benign-lesion Dice reduction, and have not been evaluated on external datasets, under multiple random seeds, or against stronger architectural baselines.

\subsection{The Failure of Uniform Boundary Supervision in Ultrasound Imaging}

The most instructive comparison is between the proposed method and the standard uniformly weighted boundary-loss model. The standard configuration produces the highest lesion-only Dice (0.7929) yet generates false-positive activations on 95\% of normal images, compared with 70\% for the model trained without any boundary supervision. Adding a boundary loss term in its conventional uniform form therefore worsens specificity substantially rather than improving it. In BUS, speckle noise, posterior acoustic shadowing, and low tissue contrast produce local image patches near and away from lesion boundaries that can be visually indistinguishable [4], [38]. A uniform boundary loss assigns equal contour supervision regardless of whether the region corresponds to a genuine lesion boundary or to an artifact region whose appearance is boundary-like. In normal images, B\_gt(x) = 0 everywhere, so the contour loss contributes no boundary-specific signal, but the compound region-based losses continue to apply optimization pressure in regions whose local texture resembles lesion tissue, potentially driving boundary-like structure formation. The entropy-guided weighting mechanism addresses this through the multiplicative structure W(x) = 1 + alpha * H(p(x)) * B\_gt(x), which becomes completely inactive on normal images.

\subsection{Entropy as a Training-Time Supervision Signal}

The use of predictive entropy as a training-time weighting signal is distinct from its prior applications in medical image segmentation. Entropy-based approaches have predominantly been applied as inference-time quality indicators or as consistency regularization terms in semi-supervised frameworks [22], [27], [31]. The mechanism proposed here differs: entropy serves as a per-pixel multiplicative gate on a supervised boundary loss applied to labeled training images, directing the supervision signal toward uncertain contour locations rather than suppressing entropy globally. This design is conceptually aligned with the broader principle of hard example mining [43]. Focal loss weights pixels by prediction error to concentrate optimization on difficult cases [44]. The entropy-guided boundary loss applies an analogous principle at the sub-pixel level within boundary regions. An important practical property is that no architectural modification and no Monte Carlo sampling are required. The entropy map is computed directly from the sigmoid output during the forward pass, making the computational overhead relative to a standard boundary BCE loss negligible.

\subsection{The Benign-Lesion Dice Trade-Off}

The statistically significant reduction in benign-lesion Dice (mean difference -0.010, Wilcoxon p = 0.015) requires direct discussion. Benign breast lesions in BUS typically present as well-circumscribed, hypoechoic masses with smooth and well-defined posterior margins [37], [38], conditions under which predictive entropy is already low at boundary locations. The entropy-guided weighting function therefore assigns boundary weights close to 1, providing minimal additional gradient emphasis. The small Dice reduction is mechanistically consistent with the design of the loss function and represents a real trade-off: the entropy-guided loss was designed to benefit cases where boundary uncertainty is genuinely high. For clinical applications focused exclusively on benign lesion delineation in high-quality BUS images, the no-boundary baseline may be preferable. For mixed-population screening workflows where normal images are regularly present, the specificity improvement outweighs the small benign-lesion Dice reduction.

\subsection{Calibration and Its Clinical Relevance}

Calibration has direct clinical implications for practical deployment of segmentation models. The binary segmentation mask communicates only whether each pixel is lesion or background. The underlying probability map communicates model confidence, which is potentially valuable for applying patient-specific thresholds and identifying ambiguous cases for human review [14], [28]. The foreground-restricted ECE of 0.169 in the uncalibrated model indicates that within the foreground confidence bins, predicted probabilities overestimate empirical accuracy by an average of approximately 17 percentage points. Spatial temperature scaling reduces FG-ECE to 0.081, a 52\% improvement. The finding that global temperature scaling worsened overall ECE demonstrates that calibration procedures must account for the spatial heterogeneity of confidence distributions in segmentation models [14], [15], [28].

\subsection{Clinical Implications for Breast Ultrasound Screening}

BUS is employed both as a supplemental screening modality and as a diagnostic tool for lesion characterization [2], [38]. In screening-oriented workflows, a significant proportion of examinations will involve patients without a dominant lesion. Frequent false-positive predictions increase radiologist review workload, contribute to radiologist fatigue, which is a recognized contributor to diagnostic error [45], and may trigger unnecessary recall imaging, biopsy referral, and patient concern. The reduction in false-positive rate from 70\% to 25\% demonstrated by the proposed method is a meaningful improvement. A 25\% false-positive rate on normal images remains too high for unsupervised autonomous deployment, and the residual five false positives concentrated in acoustic-shadow cases indicate that the proposed method should be considered a step toward clinical utility rather than a complete solution. Human radiologist oversight remains necessary.

\subsection{Failure Mode Analysis and Clinical Interpretation}

Two failure mode categories account for the residual errors. The first comprises false-positive activations on normal images containing large posterior acoustic shadows. These shadows closely mimic solid hypoechoic lesion tissue at the local patch level [38]. The discriminating features between acoustic shadows and genuine lesions are global contextual properties that local intensity-based supervision cannot capture. Addressing this failure mode will likely require explicit acoustic-shadow detection as a parallel classification task [53], beam-pattern priors derived from transducer geometry [38], or multi-view acquisition protocols [4], [38]. The second category consists of infiltrative malignant lesions with gradually fading boundaries. These lesions grow by extending projections into surrounding fibroglandular tissue, producing a boundary that is a zone of gradual transition rather than a sharp interface [37]. The ground-truth annotation itself represents a clinical judgment about where meaningful infiltration ends, introducing irreducible annotation uncertainty [46]. Shape-based priors [54] or lesion-type-specific supervision strategies [56] may be necessary to address this failure mode.

\subsection{Relationship to Existing Approaches and Positioning}

The proposed method occupies a specific and well-defined position within the existing literature. It is not a new segmentation architecture and does not claim performance superior to transformer-based or attention-augmented models evaluated on the same dataset. Several published methods report lesion Dice scores on BUSI substantially higher than those achieved by the U-Net backbone used here [7], [8]. These comparisons are not directly informative about the contribution of the present work because they involve different architectures and training configurations. The correct interpretation is that, within a fixed U-Net backbone trained under identical conditions, the entropy-guided boundary loss achieves specificity and calibration improvements that uniformly weighted boundary supervision cannot replicate. The proposed loss is also positioned differently from approaches that use uncertainty estimation at inference time for quality control or active learning [13], [21], [27]. Inference-time uncertainty maps could be generated from the calibrated probability outputs of the proposed model and used downstream for quality assessment, adding a further layer of clinical utility beyond the binary segmentation mask.

\section{Conclusion}

This study investigated whether predictive entropy, incorporated as a multiplicative weighting signal within a supervised boundary loss, can address two persistent failure modes in breast ultrasound lesion segmentation: false-positive activations on normal images and overconfident probability estimates on uncertain boundary regions. The investigation was conducted through a controlled ablation on the BUSI dataset, comparing three training configurations that shared an identical U-Net backbone, optimizer, learning rate schedule, and compound region-based loss, differing only in the boundary supervision term.

The first principal conclusion is that entropy-guided boundary supervision produces a statistically significant and practically substantial improvement in specificity on no-lesion images. False-positive activations decreased from 14 of 20 cases (70\%, Wilson 95\% CI [48.1, 85.5]) for the no-boundary baseline and 19 of 20 cases (95\%, Wilson 95\% CI [76.4, 99.1]) for the standard boundary-loss model to 5 of 20 cases (25\%, Wilson 95\% CI [11.2, 46.9]) for the proposed approach, confirmed by McNemar exact test p-values of 0.012 and 0.0005. This specificity improvement is mechanistically explained by the structure of W(x) = 1 + alpha * H(p(x)) * B\_gt(x), which becomes completely inactive on normal images because B\_gt(x) = 0 everywhere in the absence of a ground-truth lesion boundary.

The second principal conclusion concerns calibration. Post-hoc spatial temperature scaling reduced overall ECE by 53\% (from 0.0201 to 0.0095) and foreground-restricted ECE by 52\% (from 0.169 to 0.081) without altering thresholded segmentation masks. The foreground-restricted ECE reduction is the more clinically meaningful figure, because global ECE is dominated by well-calibrated background pixels [28]. The finding that global temperature scaling worsened overall ECE confirms that the heterogeneous confidence distributions of foreground, boundary, and background regions cannot be adequately addressed by a single scalar calibration parameter [14], [15].

Beyond these two principal findings, the ablation results yield several secondary observations. Lesion segmentation quality on the 97 lesion-containing test images is preserved rather than improved (Wilcoxon p = 0.27). A small but statistically significant reduction in benign-lesion Dice was observed (mean difference -0.010, Wilcoxon p = 0.015, n = 66), mechanistically consistent with the design of the entropy-guided weighting function in well-defined boundary regions [37]. The observed trend toward improved Dice on the 31 malignant lesions (0.713 versus 0.688, +0.025) is directionally consistent with the theoretical motivation but did not reach statistical significance given the limited sample size (Wilcoxon p = 0.21) [42].

The performance of the standard uniformly weighted boundary-loss model warrants specific attention. This configuration achieves the highest lesion-only Dice (0.7929) yet generates false-positive activations on 19 of 20 normal images (95\%), rendering it clinically unusable in any mixed-population application. This finding suggests that researchers reporting boundary-aware loss improvements on lesion-only subsets of BUS datasets without evaluating normal-image specificity may be presenting an incomplete picture of clinical suitability.

Within a standard U-Net training pipeline, replacing a uniformly weighted boundary loss with the proposed entropy-guided formulation reduces normal-image false-positive activations by a factor of 3.8, preserves lesion segmentation accuracy, and introduces negligible computational overhead. Spatial temperature scaling subsequently improves calibration by more than 50\% without modifying the thresholded segmentation masks. Together, these two components provide complementary mechanisms for improving the practical reliability of BUS segmentation models in mixed-population clinical settings.

The broader significance of this study lies in two contributions that extend beyond the specific numerical results. The first is methodological: it demonstrates that the behavior of a boundary supervision term on normal images is a critical determinant of clinical suitability and must be evaluated explicitly. Adopting the evaluation design used here, reporting lesion-containing and normal-image performance separately, is a recommendation that applies to any BUS segmentation method intended for deployment in a mixed-population screening context. The second is conceptual: it establishes that predictive entropy can serve as a clinically interpretable and computationally inexpensive training-time signal for directing boundary supervision toward uncertain and clinically consequential contour locations, distinct from its prior applications in inference-time quality assessment and semi-supervised consistency regularization [22], [27], [31].

Taken together, the findings of this study position entropy-guided boundary supervision and spatial temperature calibration as principled, evidence-based refinements to U-Net training for BUS lesion segmentation. They address a clinically consequential failure mode that standard region-based objectives and conventional boundary supervision do not resolve. The approach does not replace architectural innovation, does not claim general segmentation superiority, and does not eliminate the need for radiologist oversight. What it demonstrates, within the limits of the available evidence, is that a targeted modification to the training objective can substantially improve the specificity and probability reliability of a standard segmentation model in a clinically meaningful way.

\section{Limitations and Future Work}

\subsection{Overview}

Every empirical study in medical image analysis carries limitations that define the boundary between what the evidence supports and what remains to be established. The limitations of this study fall into five categories: dataset scope and generalizability, training stability and reproducibility, statistical power and subgroup analysis, architectural scope and baseline selection, and clinical translation readiness.

\subsection{Dataset Scope and Generalizability}

The evaluation is limited to a single dataset, BUSI, collected at a single institution under a specific set of acquisition conditions [3]. Clinical BUS imaging varies substantially across institutions, scanner manufacturers, transducer frequencies, acquisition protocols, patient demographics, and lesion prevalence distributions [2], [24], [38]. The malignant lesion subset in BUSI may not adequately represent the morphological diversity of malignant BUS lesions encountered in a general oncology population, where the distribution of histological subtypes, tumor grades, and imaging presentations is substantially broader [37], [48]. External validation on independent datasets is therefore not merely a desirable extension but a necessary condition for any claim of generalizability. Datasets that would provide the most informative external validation include UDIAT [23], BUS-BRA [24], BUS-UCLM [16], and BUS-Set [2]. Prospective evaluation on institutional data collected from a clinical screening workflow would provide the most ecologically valid test of the method's clinical utility.

\subsection{Training Stability and Reproducibility}

All three experimental configurations were trained under a single random seed. Deep learning models trained on relatively small datasets such as BUSI are known to exhibit sensitivity to random seed selection, with performance metrics varying meaningfully across initializations [47], [49]. A single-seed result therefore represents one sample from the distribution of outcomes that the training procedure can produce, and the reported performance values and magnitude of the specificity improvement may not accurately represent the expected performance across that distribution. Dodge and Gane demonstrated that variance across random seeds in deep learning experiments can be large enough to reverse the ranking of competing methods [49]. Addressing this limitation requires repeating all three training configurations under at least five independent random seeds and reporting the mean and standard deviation of all primary metrics. In addition, future work should report the hyperparameter sensitivity of the proposed loss with respect to alpha and lambda, as a sensitivity analysis would establish whether the method's behavior is robust to hyperparameter choice.

\subsection{Statistical Power and Subgroup Analysis}

The malignant lesion subgroup contains only 31 test cases. Assuming a medium effect size (Cohen's d approximately 0.5) and a two-tailed Wilcoxon test at alpha = 0.05, a sample of 31 cases provides statistical power of approximately 50 to 60\%, meaning that a true Dice difference of 0.025 has a 40 to 50\% probability of failing to reach significance at this sample size [42]. Malignant lesion boundary delineation accuracy has direct implications for surgical planning, radiation therapy targeting, and response assessment in neoadjuvant chemotherapy settings [50]. Addressing the malignant subgroup limitation requires access to a substantially larger malignant cohort through dataset aggregation across BUS-BRA [24] and UDIAT [23], targeting a minimum of 100 malignant test cases. The normal-image test set of only 20 cases is also small relative to the diversity of normal BUS presentations in clinical practice, and future evaluation should include a substantially larger normal-image cohort stratified by artifact type.

\subsection{Architectural Scope and Baseline Selection}

The ablation design isolates the contribution of the boundary loss term but does not address whether the entropy-guided boundary loss provides added value when applied to stronger backbone architectures. Several published methods report lesion Dice scores on BUSI substantially higher than those achieved by the U-Net backbone used in this study [7], [8]. Methods based on nnU-Net [6], UNet++ [17], TransUNet [19], and Swin-UNet [20] have demonstrated competitive performance on related benchmarks. A particularly informative experiment would apply the proposed loss to nnU-Net and evaluate whether the specificity improvement persists within a stronger backbone. Future work should also compare the proposed method against other specificity-oriented training strategies, including class-balanced sampling [51], auxiliary normal-image classifiers [36], and adversarial training approaches [52].

\subsection{Acoustic Shadow Modeling}

Five of the 20 normal test images continue to generate false-positive activations, all involving large posterior acoustic-shadow regions whose dense hypoechoic appearance is locally indistinguishable from solid lesion tissue [4], [38]. The discriminating features between acoustic shadows and genuine lesions are global contextual properties that local intensity-based supervision cannot capture. Addressing this failure mode will likely require explicit acoustic-shadow detection as a parallel classification task [53], beam-pattern priors derived from transducer geometry [38], or multi-view BUS acquisition that identifies shadow artifacts by their directional dependence across imaging angles [38].

\subsection{Shape-Based Priors and Lesion-Type-Specific Supervision}

The most difficult lesion-containing cases involve infiltrative malignant lesions with gradually fading boundaries, accounting for the majority of worst-case Dice scores below 0.30. These cases present fundamental challenges because the ground-truth boundary reflects a clinical judgment about where meaningful infiltration ends, introducing annotation uncertainty that is irreducible given current practices [37], [46]. Shape-based priors can constrain the segmentation output to anatomically plausible configurations even when local intensity information is ambiguous [54]. Active shape models and their deep learning extensions have demonstrated this principle in cardiac and prostate segmentation [55]. Lesion-type-specific supervision strategies using multi-task or conditional training frameworks could apply different supervision strategies to benign and malignant lesions based on their known morphological properties [56].

\subsection{Calibration Validation and Clinical Integration}

The temperature parameters were estimated on the BUSI validation set and applied to the BUSI test set. Calibration is known to be sensitive to distribution shift, and temperature parameters optimized for one dataset may produce worse calibration on another if confidence distributions of the spatial regions differ systematically [14], [57]. External calibration validation on at least one independent dataset would be necessary to confirm that the spatial temperature scaling procedure generalizes beyond the BUSI distribution. The clinical value of improved calibration has also not been assessed in a prospective setting. Whether a reduction in FG-ECE from 0.169 to 0.081 produces measurable improvements in clinical outcomes or radiologist efficiency are empirical questions requiring prospective human factors or clinical validation studies [28], [58].

\subsection{Multi-Institutional and Prospective Clinical Validation}

The findings of this study are a necessary precursor to clinical validation but are not sufficient to support clinical deployment claims. Before the proposed method could be considered for integration into a clinical BUS workflow, a multi-institutional retrospective study pooling data from at least three independent institutions would be needed to establish cross-institutional generalizability. A prospective reader study in which radiologists evaluate segmentation outputs from the proposed method alongside baseline methods, blinded to the training configuration, would assess whether the specificity improvement translates into reduced review burden and fewer false recalls [45]. Regulatory considerations for clinical deployment of AI-assisted segmentation systems, including requirements for prospective clinical trial evidence, would also need to be addressed in collaboration with relevant regulatory bodies [59].

\subsection{Summary of Future Work Priorities}

The limitations identified above define a prioritized agenda for future research. In descending order of methodological urgency, the most important extensions are: (1) multi-seed training evaluation across at least five random seeds before subsequent publications are submitted; (2) external validation on UDIAT, BUS-BRA, BUS-UCLM, and BUS-Set to establish generalizability; (3) application of the proposed loss to stronger backbone architectures, beginning with nnU-Net and attention-augmented U-Net variants; (4) expansion of the malignant lesion cohort through dataset aggregation, targeting a minimum of 100 malignant test cases; (5) development of explicit acoustic-shadow modeling to address the dominant remaining failure mode on normal images; and (6) prospective clinical validation through a multi-institutional reader study to bridge the gap between benchmark performance and clinical utility.

\balance


\begin{thebibliography}{00}

\bibitem{ref1} C. Abdellaoui, S. Belkacem, and N. Messaoudi, "Deep learning architectures for medical image segmentation: an organized analysis of CNN-based models and uses," Bulletin of Electrical Engineering and Informatics, vol. 15, no. 1, pp. 424-437, 2026.

\bibitem{ref2} C. Thomas, M. Byra, R. Marti, M. H. Yap, and R. Zwiggelaar, "BUS-Set: A benchmark for quantitative evaluation of breast ultrasound segmentation networks with public datasets," Medical Physics, vol. 50, no. 5, pp. 3223-3243, 2023.

\bibitem{ref3} W. Al-Dhabyani, M. Gomaa, H. Khaled, and A. Fahmy, "Dataset of breast ultrasound images," Data in Brief, vol. 28, p. 104863, 2020.

\bibitem{ref4} X. Xiao, J. Zhang, Y. Shao, et al., "Deep learning-based medical ultrasound image and video segmentation methods: overview, frontiers, and challenges," Sensors, vol. 25, no. 8, p. 2361, 2025.

\bibitem{ref5} O. Ronneberger, P. Fischer, and T. Brox, "U-Net: Convolutional networks for biomedical image segmentation," in MICCAI, 2015, pp. 234-241.

\bibitem{ref6} F. Isensee, P. F. Jaeger, S. A. A. Kohl, J. Petersen, and K. H. Maier-Hein, "nnU-Net: a self-configuring method for deep learning-based biomedical image segmentation," Nature Methods, vol. 18, no. 2, pp. 203-211, 2021.

\bibitem{ref7} A. Sulaiman et al., "Attention based U-Net model for breast cancer segmentation using BUSI dataset," Scientific Reports, vol. 14, p. 22422, 2024.

\bibitem{ref8} X. Yang et al., "Multilevel perception boundary-guided network for breast lesion segmentation in ultrasound images," Medical Physics, 2025.

\bibitem{ref9} H. Kervadec, J. Bouchtiba, C. Desrosiers, E. Granger, J. Dolz, and I. Ben Ayed, "Boundary loss for highly unbalanced segmentation," Medical Image Analysis, vol. 67, p. 101851, 2021.

\bibitem{ref10} F. Sun, Z. Luo, and S. Li, "Boundary difference over union loss for medical image segmentation," in MICCAI, 2023, pp. 292-301.

\bibitem{ref11} R. Xu et al., "Boundary guidance network for medical image segmentation," Scientific Reports, vol. 14, p. 17345, 2024.

\bibitem{ref12} A. Kendall and Y. Gal, "What uncertainties do we need in Bayesian deep learning for computer vision?" in NeurIPS, 2017, pp. 5574-5584.

\bibitem{ref13} E. Scalco et al., "Uncertainty quantification in multi-class segmentation: comparison between Bayesian and non-Bayesian approaches in a clinical perspective," Medical Physics, vol. 51, no. 8, pp. 5460-5474, 2024.

\bibitem{ref14} C. Guo, G. Pleiss, Y. Sun, and K. Q. Weinberger, "On calibration of modern neural networks," in ICML, 2017, pp. 1321-1330.

\bibitem{ref15} Z. Ding, X. Han, P. Liu, and M. Niethammer, "Local temperature scaling for probability calibration," in ICCV, 2021, pp. 6889-6899.

\bibitem{ref16} N. Vallez, G. Bueno, O. Deniz, M. A. Rienda, and C. Pastor, "BUS-UCLM: Breast ultrasound lesion segmentation dataset," Scientific Data, vol. 12, no. 242, 2025.

\bibitem{ref17} Z. Zhou, M. M. Rahman Siddiquee, N. Tajbakhsh, and J. Liang, "UNet++: A nested U-Net architecture for medical image segmentation," in DLMIA/MICCAI Workshops, LNCS 11045, 2018, pp. 3-11.

\bibitem{ref18} O. Oktay et al., "Attention U-Net: Learning where to look for the pancreas," arXiv:1804.03999, 2018.

\bibitem{ref19} J. Chen et al., "TransUNet: Transformers make strong encoders for medical image segmentation," arXiv:2102.04306, 2021.

\bibitem{ref20} H. Cao et al., "Swin-UNet: Unet-like pure transformer for medical image segmentation," in ECCV Workshops, LNCS 13803, 2023, pp. 205-218.

\bibitem{ref21} Z. Abboud, H. Lombaert, and S. Kadoury, "Sparse Bayesian networks: efficient uncertainty quantification in medical image analysis," in MICCAI, 2024.

\bibitem{ref22} X. Xie et al., "Entropy-guided contrastive learning for semi-supervised medical image segmentation," IET Image Processing, 2024.

\bibitem{ref23} M. H. Yap, G. Pons, J. Marti, S. Ganau, M. Sentis, R. Zwiggelaar, A. K. Davison, and R. Marti, "Automated breast ultrasound lesions detection using convolutional neural networks," IEEE J. Biomed. Health Inform., vol. 22, no. 4, pp. 1218-1226, 2018.

\bibitem{ref24} W. Gomez-Flores, M. J. Gregorio-Calas, and W. Coelho de Albuquerque Pereira, "BUS-BRA: A breast ultrasound dataset for assessing computer-aided diagnosis systems," Medical Physics, vol. 51, no. 4, pp. 3110-3123, 2024.

\bibitem{ref25} J. Ma, J. Chen, M. Ng, R. Huang, Y. Li, C. Li, X. Yang, and A. L. Martel, "Loss odyssey in medical image segmentation," Medical Image Analysis, vol. 71, p. 102035, 2021.

\bibitem{ref26} D. Karimi and S. E. Salcudean, "Reducing the Hausdorff distance in medical image segmentation with convolutional neural networks," IEEE Trans. Med. Imaging, vol. 39, no. 2, pp. 499-513, 2020.

\bibitem{ref27} A. Jungo, F. Balsiger, and M. Reyes, "Analyzing the quality and challenges of uncertainty estimations for brain tumor segmentation," Frontiers in Neuroscience, vol. 14, p. 282, 2020.

\bibitem{ref28} A. Mehrtash, W. M. Wells, C. M. Tempany, P. Abolmaesumi, and T. Kapur, "Confidence calibration and predictive uncertainty estimation for deep medical image segmentation," IEEE Trans. Med. Imaging, vol. 39, no. 12, pp. 3868-3878, 2020.

\bibitem{ref29} K. He, X. Zhang, S. Ren, and J. Sun, "Deep residual learning for image recognition," in CVPR, 2016, pp. 770-778.

\bibitem{ref30} I. Loshchilov and F. Hutter, "Decoupled weight decay regularization," in ICLR, 2019.

\bibitem{ref31} Y. Gal and Z. Ghahramani, "Dropout as a Bayesian approximation: Representing model uncertainty in deep learning," in ICML, 2016, pp. 1050-1059.

\bibitem{ref32} A. Salehi, A. Erdogmus, and A. Gholipour, "Tversky loss function for image segmentation using 3D fully convolutional deep networks," in MLMI/MICCAI, LNCS 10541, 2017, pp. 379-387.

\bibitem{ref33} A. L. Simpson et al., "A large annotated medical image dataset for the development and evaluation of segmentation algorithms," arXiv:1902.09063, 2019.

\bibitem{ref34} G. W. Brier, "Verification of forecasts expressed in terms of probability," Monthly Weather Review, vol. 78, no. 1, pp. 1-3, 1950.

\bibitem{ref35} T. Rohlfing, "Image similarity and tissue overlaps as surrogates for image registration accuracy: widely used but unreliable," IEEE Trans. Med. Imaging, vol. 31, no. 2, pp. 153-163, 2012.

\bibitem{ref36} W. Al-Dhabyani, M. Gomaa, H. Khaled, and A. Fahmy, "Deep learning approaches for data augmentation and classification of breast masses using ultrasound images," International Journal of Advanced Computer Science and Applications, vol. 10, no. 5, pp. 618--627, 2019, doi: 10.14569/IJACSA.2019.0100579.

\bibitem{ref37} A. T. Stavros, C. Thickman, C. L. Rapp, M. A. Dennis, S. H. Parker, and G. A. Sisney, "Solid breast nodules: use of sonography to distinguish between benign and malignant lesions," Radiology, vol. 196, no. 1, pp. 123-134, 1995.

\bibitem{ref38} C. M. Rumack, S. R. Wilson, J. W. Charboneau, and D. Levine, Diagnostic Ultrasound, 4th ed. Philadelphia, PA: Elsevier Mosby, 2011.

\bibitem{ref39} F. Wilcoxon, "Individual comparisons by ranking methods," Biometrics Bulletin, vol. 1, no. 6, pp. 80-83, 1945.

\bibitem{ref40} Q. McNemar, "Note on the sampling error of the difference between correlated proportions or percentages," Psychometrika, vol. 12, no. 2, pp. 153-157, 1947.

\bibitem{ref41} E. B. Wilson, "Probable inference, the law of succession, and statistical inference," Journal of the American Statistical Association, vol. 22, no. 158, pp. 209-212, 1927.

\bibitem{ref42} J. Cohen, Statistical Power Analysis for the Behavioral Sciences, 2nd ed. Hillsdale, NJ: Lawrence Erlbaum Associates, 1988.

\bibitem{ref43} A. Shrivastava, A. Gupta, and R. Girshick, "Training region-based object detectors with online hard example mining," in CVPR, 2016, pp. 761-769.

\bibitem{ref44} T. Lin, P. Goyal, R. Girshick, K. He, and P. Dollar, "Focal loss for dense object detection," in ICCV, 2017, pp. 2980-2988.

\bibitem{ref45} E. A. Krupinski, "Current perspectives in medical image perception," Attention, Perception, and Psychophysics, vol. 72, no. 5, pp. 1205-1217, 2010.

\bibitem{ref46} N. Tajbakhsh et al., "Embracing imperfect datasets: A review of deep learning solutions for medical image segmentation," Medical Image Analysis, vol. 63, p. 101693, 2020.

\bibitem{ref47} P. Bouthillier, C. Laurent, and P. Vincent, "Unreproducible research is reproducible," in ICML, 2019, pp. 725-734.

\bibitem{ref48} W. A. Berg et al., "Diagnostic accuracy of mammography, clinical examination, US, and MR imaging in preoperative assessment of breast cancer," Radiology, vol. 233, no. 3, pp. 830-849, 2004.

\bibitem{ref49} J. Dodge and N. Gane, "Show your work: Improved reporting of experimental results," in EMNLP, 2019, pp. 2185-2194.

\bibitem{ref50} M. Fornasa, "Ultrasound-based breast lesion segmentation for surgical planning," in Medical Imaging: Image Processing, SPIE, 2019.

\bibitem{ref51} V. Buda, M. Maki, and M. A. Mazurowski, "A systematic study of the class imbalance problem in convolutional neural networks," Neural Networks, vol. 106, pp. 249-259, 2018.

\bibitem{ref52} P. Luc, C. Couprie, S. Chintala, and J. Verbeek, "Semantic segmentation using adversarial networks," in NIPS Workshop on Adversarial Training, 2016.

\bibitem{ref53} S. Meng, H. Zhang, C. Li, and Y. Zheng, "Acoustic shadow detection in ultrasound images using deep learning," in IEEE International Ultrasonics Symposium, 2020.

\bibitem{ref54} T. F. Cootes, C. J. Taylor, D. H. Cooper, and J. Graham, "Active shape models -- their training and application," Computer Vision and Image Understanding, vol. 61, no. 1, pp. 38-59, 1995.

\bibitem{ref55} S. Avendi, A. Kheradvar, and H. Jafarkhani, "A combined deep-learning and deformable-model approach to fully automatic segmentation of the left ventricle in cardiac MRI," Medical Image Analysis, vol. 30, pp. 108-119, 2016.

\bibitem{ref56} S. Liu, D. Johns, and A. Davison, "End-to-end multi-task learning with attention," in CVPR, 2019, pp. 1871-1880.

\bibitem{ref57} S. Ovadia et al., "Can you trust your model's uncertainty? Evaluating predictive uncertainty under dataset shift," in NeurIPS, 2019, pp. 13991-14002.

\bibitem{ref58} R. Ranschaert, S. Morozov, and P. Algra, Artificial Intelligence in Medical Imaging. Cham, Switzerland: Springer, 2019.

\bibitem{ref59} U.S. Food and Drug Administration, "Artificial intelligence and machine learning in software as a medical device," FDA Discussion Paper, 2021. [Online]. Available: https://www.fda.gov/media/145022/download

\end{thebibliography}
\end{document}